\documentclass[prl,twocolumn,a4paper,floatfix]{revtex4}
\usepackage{graphics}
\usepackage{bbm}
\usepackage{amsmath}

\def\empile#1\over#2{\mathrel{\mathop{\kern 0pt#1}\limits_{#2}}}
\newcommand{\slvarepsilon}{\raise.15ex\hbox{$/$}\kern-.53em\hbox{$\varepsilon$}}
\newcommand{\slL}{\raise.15ex\hbox{$/$}\kern-.53em\hbox{$L$}}
\newcommand{\slP}{\raise.15ex\hbox{$/$}\kern-.53em\hbox{$P$}}
\newcommand{\slp}{\raise.1ex\hbox{$/$}\kern-.63em\hbox{$p$}}
\newcommand{\slq}{\raise.1ex\hbox{$/$}\kern-.53em\hbox{$q$}}
\newcommand{\slv}{\raise.1ex\hbox{$/$}\kern-.63em\hbox{$v$}}
\newcommand{\slR}{\raise.15ex\hbox{$/$}\kern-.53em\hbox{$R$}}
\newcommand{\slQ}{\raise.15ex\hbox{$/$}\kern-.53em\hbox{$Q$}}
\newcommand{\slK}{\raise.15ex\hbox{$/$}\kern-.53em\hbox{$K$}}
\newcommand{\slk}{\raise.15ex\hbox{$/$}\kern-.53em\hbox{$k$}}
\newcommand{\slSigma}{\raise.15ex\hbox{$/$}\kern-.53em\hbox{$\Sigma$}}
\newcommand{\slcalP}{\raise.15ex\hbox{$/$}\kern-.63em\hbox{$\cal P$}}
\newcommand{\slA}{\raise.15ex\hbox{$/$}\kern-.73em\hbox{$A$}}
\newcommand{\slbfA}{\raise.15ex\hbox{$/$}\kern-.73em\hbox{${\imb A}$}}
\newcommand{\slpartial}{\raise.15ex\hbox{$/$}\kern-.53em\hbox{$\partial$}}
\newcommand{\sla}{\raise.15ex\hbox{$/$}\kern-.53em\hbox{$a$}}
\newcommand{\slb}{\raise.15ex\hbox{$/$}\kern-.53em\hbox{$b$}}
\newcommand{\slc}{\raise.15ex\hbox{$/$}\kern-.53em\hbox{$c$}}
\newcommand{\slC}{\raise.15ex\hbox{$/$}\kern-.63em\hbox{$C$}}

\def\p{{\boldsymbol p}}
\def\q{{\boldsymbol q}}
\def\l{{\boldsymbol l}}
\def\k{{\boldsymbol k}}

\def\x{{\boldsymbol x}}

\def\y{{\boldsymbol y}}

\def\wt{\widetilde}

\begin{document}

\title{Quantitative study of the violation of $k_\perp$-factorization\\
 in hadroproduction of quarks at collider energies}

\author{Hirotsugu Fujii$^{\rm a}$,
        Fran\c cois Gelis$^{\rm b}$
        and
        Raju Venugopalan$^{\rm c,d}$
        }

\affiliation{%
 a\ Institute of Physics, University of Tokyo, Komaba, 
 Tokyo 153-8902, Japan.\\ 
 b\ CEA/DSM/SPhT, 91191 Gif-sur-Yvette cedex, France.\\
 c\ Physics Department, Brookhaven National Laboratory, Upton,
 N.Y. 11973, U.S.A.\\
 d\ ECT$\star$, Villa Tambosi, Strada delle Tabarelle 286, I-38050, Villazzano(TN), Italy.\\
}

\date{\today}

\begin{abstract}
We demonstrate the violation of $k_\perp$-factorization for quark production in 
high energy hadronic collisions.  This violation is quantified in the Color Glass Condensate framework
and studied as a function of the quark mass, the quark transverse momentum, and the saturation 
scale $Q_s$, which is a measure of large parton densities. At $x$ values where parton densities are large but leading twist shadowing effects 
are still small, violations of $k_\perp$-factorization can be significant -- especially for lighter quarks. At very small $x$, where
leading twist shadowing is large, we show that violations of $k_\perp$-factorization are relatively weaker. 
\end{abstract}


\maketitle

The $k_\perp$-factorization formalism~\cite{ktfact}, devised
originally for heavy quark production in hadronic collisions,
systematically resums powers of $\alpha_s \ln(s/q_\perp^2)$ in
perturbative QCD. These contributions are important at high energies
when the transverse momentum $q_\perp$ of the final state is much
smaller than the center of mass energy $\sqrt{s}$. In the
$k_\perp$-factorization framework, the quark production cross-section
is expressed as the convolution of a hard matrix element and
distribution functions from each of the two hadrons. These
``unintegrated'' gluon distributions depend on the longitudinal
momentum fraction $x$ and the transverse momentum $\k_\perp$ of the
gluon taking part in the hard scattering. When integrated over all
transverse momenta up to some hard scale $M^2$, these distributions
give the more familiar gluon distribution $x\,G(x,M^2)$.

Even though logarithms due to gluon branchings are resummed in the framework of refs.~\cite{ktfact}, 
only one hard gluon from each projectile participates in the reaction. It is interesting to consider whether the
$k_\perp$-factorization framework can be extended beyond the single
hard scattering (leading twist) case. The
simplest experiments for studying hard multiple scattering (higher twist) effects are those where 
one of the projectiles is dilute and the other is dense. These include a) p-p
collisions in the forward/backward fragmentation region at the Relativistic Heavy Ion Collider (RHIC) and 
the Large Hadron Collider (LHC), where large
$x$'s in one projectile and small $x$'s in the other are probed, and
b) proton or Deuteron collisions off large nuclei at RHIC and LHC. Understanding the validity of 
$k_\perp$ factorization in this multiple scattering regime 
is important for a quantitative understanding of final states at RHIC and LHC.

The Color Glass Condensate (CGC) \cite{CGC}, wherein the two
projectiles are sources of classical color fields, is a
powerful framework to study multiple scattering effects.  The
leading twist results of refs. \cite{ktfact} for quark pair production
are easily recovered in this formalism \cite{GelisV1} keeping only terms that are of 
lowest order in the sources.  $k_\perp$-factorization was studied in the forward p-p/p-A case, where one keeps the lowest
order in one of the sources and all orders in the other source -- this
power counting is discussed further after
eq.~(\ref{eq:corr-nucleus}). In refs.~\cite{gluons1,BlaizGV1}, this formalism was
used to obtain the cross-section for gluon production in p-A
collisions, which was found to be $k_\perp$-factorizable
in~\cite{BlaizGV1,gluons2}.  Quark production in p-A collisions was
computed in the CGC formalism in~\cite{BlaizGV2,Tuchin}. It was shown
explicitly in ~\cite{BlaizGV2} that both quark pair production and
single inclusive quark cross-sections are not 
$k_\perp$-factorizable~\footnote{The violation of
$k_\perp$-factorization has been studied in a different framework. See
ref.~\cite{NikolSZ1} and references therein.}.  In this note, we quantify, for the 
first time, the magnitude of the violation of $k_\perp$ factorization. 

The generalization of the leading twist formula of
refs.~\cite{ktfact} for the inclusive quark production
cross-section in p-A collisions gives~\cite{BlaizGV2},
\begin{eqnarray}
&&\frac{d\sigma_{_Q}}{d^2\q_\perp dy_q}
=
\frac{\alpha_s^2 N}{8\pi^4 (N^2-1)}\int\frac{dp^+}{p^+}
\int_{\k_{1\perp},\k_{2\perp}}
\frac{1}{\k_{1\perp}^2 \k_{2\perp}^2}
\nonumber\\
&&
\times\Big\{
{\rm tr}
\Big[(\slq\!+\!m)T_{q\bar{q}}(\slp\!-\!m)
T_{q\bar{q}}^{*}\Big]
\frac{C_{_{F}}}{N}\phi_{_A}^{q,q}
(\k_{2\perp})
\nonumber\\
&&
+\int_{\k_\perp}
\!
{\rm tr}
\Big[(\slq\!+\!m)T_{q\bar{q}}(\slp\!-\!m)
T_{g}^{*}\Big]
\phi_{_A}^{q\bar{q},g}
(\k_{2\perp};\k_\perp)+{\rm h.c.}
\nonumber\\
&&\;
+{\rm tr}
\Big[(\slq\!+\!m)T_{g}(\slp\!-\!m)T_{g}^{*}\Big]
\phi_{_A}^{g,g}(\k_{2\perp})
\Big\}
\varphi_p(\k_{1\perp})\; ,
\label{eq:cross-section-q}
\end{eqnarray}
where the shorthand notation corresponds to the explicit expressions,
\begin{eqnarray}
&&T_{q\bar{q}}(\k_{1\perp},\k_{\perp})\equiv
\nonumber\\
&&\equiv
\frac{\gamma^+(\slq-\slk+m)\gamma^-(\slq-\slk-\slk_1+m)\gamma^+}
{2p^+[(\q_\perp\!-\!\k_\perp)^2\!+\!m^2]\!
+\!2q^+[(\q_\perp\!-\!\k_\perp\!-\!\k_{1\perp})^2\!+\!m^2]}
\nonumber\\
&&T_{g}(\k_{1\perp})\equiv 
\frac{\slC_{_{L}}(p+q,\k_{1\perp})}{(p+q)^2}
\; .
\label{eq:Tqqbar-Tg}
\end{eqnarray}
$C_{_L}^\mu$ is the well known Lipatov effective vertex. In
eq.~(\ref{eq:cross-section-q}), $\k_{1\perp}$ and $\k_{2\perp}$ are
respectively the transverse momenta transferred from the proton and the
nucleus, and $\p_\perp$ must be understood as
$\k_{1\perp}+\k_{2\perp}-\q_\perp$. All the relevant information about
the proton and the nucleus is encoded in the function $\varphi_p$ and
in the various $\phi_{_A}$'s respectively. These are defined as 
\begin{equation}
\varphi_p(\l_\perp)\equiv {\pi^2 R_p^2 g^2 \over l_\perp^2}\, 
\int_{\x_\perp}
e^{i\l_\perp\cdot\x_\perp}\left<\rho_1^a(0)\rho_1^a(\x_\perp)\right>\; ,
\label{eq:corr-proton}
\end{equation}
for the proton, and
\begin{eqnarray}
&&\phi_{_A}^{q,q}(\l_\perp)\equiv \frac{2\pi^2 R_{_A}^2 l_\perp^2}{g^2}
\int_{\x_\perp}
e^{i\l_\perp\cdot\x_\perp}
{\rm tr}\,\left<
{\wt U}(0)
{\wt U}^\dagger(\x_\perp)
\right>\; ,
\nonumber\\
&&\phi_{_{A}}^{g,g}(\l_{\perp})\equiv \frac{\pi^2 R_{_A}^2 l_\perp^2}{g^2 N}
\int_{\x_\perp}
e^{i\l_\perp\cdot\x_\perp}
\,{\rm tr}\,\left<U(0)U^\dagger(\x_\perp)\right>\; ,
\nonumber\\
&&\phi_{_A}^{q\bar{q},g}(\l_\perp;\k_\perp)\equiv
\frac{2\pi^2 R_{_A}^2 l_\perp^2}{g^2 N}
\int_{\x_\perp,\y_\perp}\!\!\!\!\!\!\!\!\!\!
e^{i(\k_\perp\cdot\x_\perp+(\l_\perp-\k_\perp)\cdot\y_\perp)}
\nonumber\\
&&\qquad\qquad\qquad\times
{\rm tr}
\left<
{\wt U}(\x_\perp)t^a {\wt U}^\dagger(\y_\perp) t^b U_{ba}(0)
\right>
\label{eq:corr-nucleus}
\end{eqnarray}
for the nucleus.  $R_p$ and $R_{_A}$ are the radii of the proton and
the nucleus respectively.  $U$ is a Wilson line in the adjoint
representation, and ${\wt U}$ is a Wilson line in the fundamental
representation.  The function $\varphi_p$, expressed here in terms of
the color charge density $\rho_1^a$ in the proton, is the usual
leading twist unintegrated gluon distribution of the
proton~\cite{GelisV1}. The functions $\phi_{_A}$, on the other hand,
include higher twist re-scattering corrections to all orders in
$\rho_2^a/l_\perp^2$, the ratio of the color charge density in the
nucleus divided by the square of the momentum $l_\perp^2$ transferred
from the nucleus. These correlators, and models to compute these, will
be discussed further shortly. We will only mention here that
the three $\phi_{_A}$'s obey, in full generality, the sum rules,
\begin{eqnarray}
&&\int\frac{d^2\l_\perp}{(2\pi)^2}\!\!
\frac{\phi_{_A}^{g,g}(\l_\perp)}{\l_\perp^2}
=\frac{C_{_F}}{N}\!\!
\int\frac{d^2\l_\perp}{(2\pi)^2}\!\!
\frac{\phi_{_A}^{q,q}(\l_\perp)}{\l_\perp^2}
={2\pi^2 R_{_A}^2 C_{_F}\over g^2}\; ,
\nonumber\\
&&\int\frac{d^2\k_\perp}{(2\pi)^2}\phi_{_A}^{q\bar{q},g}(\l_\perp;\k_\perp)
=\phi_{_A}^{g,g}(\l_\perp)\; .
\end{eqnarray}

Equation (\ref{eq:cross-section-q}) has another important
property. When integrated over $p^+$, the Dirac traces that
appear in this formula become independent of $q^+$, or equivalently,
of the rapidity of the produced quark. In a model 
like the McLerran-Venugopalan (MV) model \cite{MV}, where the
distributions $\varphi_p$ and $\phi_{_A}$ are independent of $x$, the
single-quark cross-section is rigorously rapidity independent. When
the distributions have an $x$ dependence, it is a priori not
legitimate to integrate the Dirac traces over $p^+$ because of the $p^+$ dependence in the $x$ variables.  However, doing so is a
good approximation if the $p^+$ dependence of the Dirac traces
is much stronger than that of the unintegrated gluon distributions. In this approximation, the integral over $p^+$ of the Dirac
traces can then be done in closed form. One of these is particularly simple and gives
\begin{equation}
\int\limits_0^{+\infty}\frac{dp^+}{p^+}{\rm tr}
\Big[(\slq\!+\!m)T_{q\bar{q}}(\slp\!-\!m) T_{q\bar{q}}^{*}\Big]=8\; .
\end{equation}
 Therefore, the sum rule obeyed by the distribution
$\phi_{_A}^{q,q}$ applies for this term and we can replace the factor
$(C_{_F}/N)\phi_{_A}^{q,q}$ by $\phi_{_A}^{g,g}$ without changing the
result. Thus the only term in eq.~(\ref{eq:cross-section-q}) that
truly breaks $k_\perp$-factorization is the term involving the 3-point
correlator $\phi_{_A}^{q\bar{q},g}$.  For the other two terms in the
cross-section, the result of the analytic integration over $p^+$ is
rather involved and will not be quoted here.

We shall now  compute the correlators in
eqs.~(\ref{eq:corr-proton}) and ~(\ref{eq:corr-nucleus}) in models which give good results in specific kinematic
regions. As discussed more extensively in ~\cite{BlaizGV1}, studies of Gold-Gold and Deuteron-Gold collisions at RHIC suggest that 
the classical MV model, where the expectation values of the Wilson line operators are computed with Gaussian source functionals,
is a good model for source distributions of large nuclei at moderately
small $x$ ($x\sim 10^{-2}$). In this kinematic region, where small $x$ quantum evolution is not dominant, multiple scattering contributions
can be interpreted entirely as higher twist effects. These die away rapidly with transverse momentum. When small $x$ quantum evolution 
becomes significant, the Gaussian distribution of sources breaks down. The evolution of the source distributions at smaller values of $x$ is 
described by a functional renormalization group (RG) equation~\cite{CGC} which remains to be solved. However, at very high parton densities, 
color charges are strongly screened allowing the application of mean field techniques analogous to the well known Random Phase Approximation.   
In this limit, the source functional is remarkably also a Gaussian, though (crucially) a non-local one~\cite{IancuIM2}.  Unlike the local MV model, 
which does not include non-trivial screening and recombination effects, one obtains leading twist shadowing from this Gaussian mean
 field model (GMF). As discussed in ~\cite{BlaizGV1}, it is conceivable that this regime might already be reached for $x\sim 10^{-3}$ in large 
 nuclei. In the absence of a full solution of the RG equations, the MV model and the GMF model correspond to two extremes of high parton 
 density behavior. We expect more quantitative results for single quark production to lie between these two extremes. 

\begin{figure}[htbp]
\begin{center}
\resizebox*{6.5cm}{!}{\includegraphics{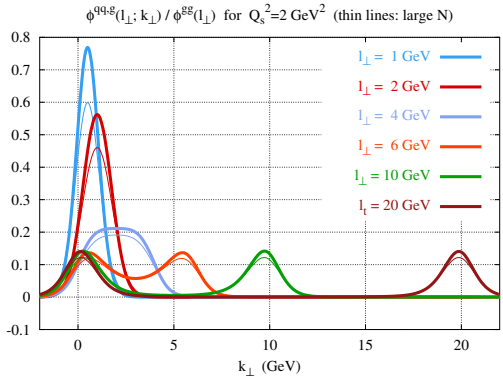}}
\end{center}
\caption{\label{fig:phi-qqg}The ratio
  $\phi_{_A}^{q\bar{q},g}(\l_\perp;\k_\perp)/
    \phi_{_A}^{g,g}(\l_\perp)$ as a function of $\k_\perp$
  (along a line parallel to $\l_\perp$ in the $\k_\perp$ plane) for
  various values of $\l_\perp$ and $Q_s^2 = 2~{\rm GeV}^2$. The thin
  lines represent the same ratio evaluated with the large $N$
  approximation for the 3-point correlator in the numerator.}
\end{figure}

In Appendix A of ref.~\cite{BlaizGV2}, we derived analytic expressions
for the correlators of Wilson lines that appear in the $\phi_{_A}$'s when the distribution of color sources in the nucleus is
a Gaussian, be they local or non-local. We shall use these results here. The 2-point correlators for these models in momentum space are easily
computed since they reduce to a 1-dimensional integral. This follows
from rotational symmetry in the transverse plane. The 3-point function
$\phi_{_A}^{q\bar{q},g}$ is more computationally intensive because
it is determined by a 3-dimensional Fourier integral. $k_\perp$-factorization requires that the 3-point
and 2-point correlators are related by~\cite{BlaizGV2}
\begin{equation}
\phi_{_A}^{q\bar{q},g}(\l_\perp;\k_\perp)
=\frac{1}{2}(2\pi)^2
\left[
\delta(\k_\perp)+\delta(\k_\perp-\l_\perp)
\right]
\phi_{_A}^{g,g}(\l_\perp)\; .
\label{eq:3p-kt-fact}
\end{equation}
This can be checked by substituting this relation in
eq.~(\ref{eq:cross-section-q}).  The ratio, for the MV model, between
the exact 3-point function and the 2-point function $\phi_{_A}^{g,g}$
as a function of $\k_\perp$ at fixed $\l_\perp$ is displayed in figure
\ref{fig:phi-qqg}.  
At large enough $\l_\perp$, this ratio indeed has
two peaks centered at $\k_\perp=0$ and $\k_\perp=\l_\perp$
respectively.  The width of these peaks is roughly of the order of the
saturation momentum in the nucleus, $Q_s$. Eventually, when $\l_\perp$
decreases below a value of order $Q_s$, the two peaks merge into a
single maximum centered at $\l_\perp/2$. This behavior of the three
point correlator suggests that $k_\perp$-factorization is a good
approximation if the produced quark is characterized by some large
momentum scale ($m$ or $\q_\perp$). Indeed, if this is the case, the
typical $\k_{2\perp}$ in eq.~(\ref{eq:cross-section-q}) is large
compared to $Q_s$, and eq.~(\ref{eq:3p-kt-fact}) is a good
approximation of the two-peak structure of the exact 3-point function.

\begin{figure}[ht]
\begin{center}
\resizebox*{6.5cm}{!}{\includegraphics{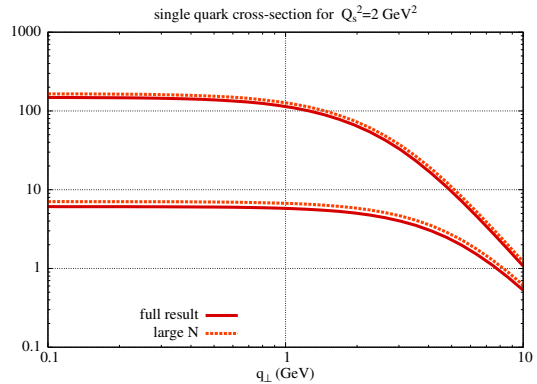}}
\end{center}
\caption{\label{fig:large-N}Comparison of the cross sections obtained
  with the exact 3-point function and with the large $N$
  approximation, for $Q_s^2=2~{\rm GeV}^2$, $m=1.5$~GeV and
  $m=4.5$~GeV.}
\end{figure}

In the large $N$ limit, the 3-point function in the MV and GMF models is a
product of two 2-point functions. 
The validity of the large N computation of the 3-point
function (represented by thin lines in figure
\ref{fig:phi-qqg}) is reasonably good. In figure \ref{fig:large-N},
we compare the quark cross-sections obtained in the
full calculation with those in the large $N$ approximation computed
for $N=3$. They are close to one another since terms neglected in the large $N$ limit are of order $1/N^2$. We shall therefore 
perform all subsequent numerical calculations in the large $N$ approximation.

\begin{figure}[htbp]
\begin{center}
\resizebox*{6.5cm}{!}{\includegraphics{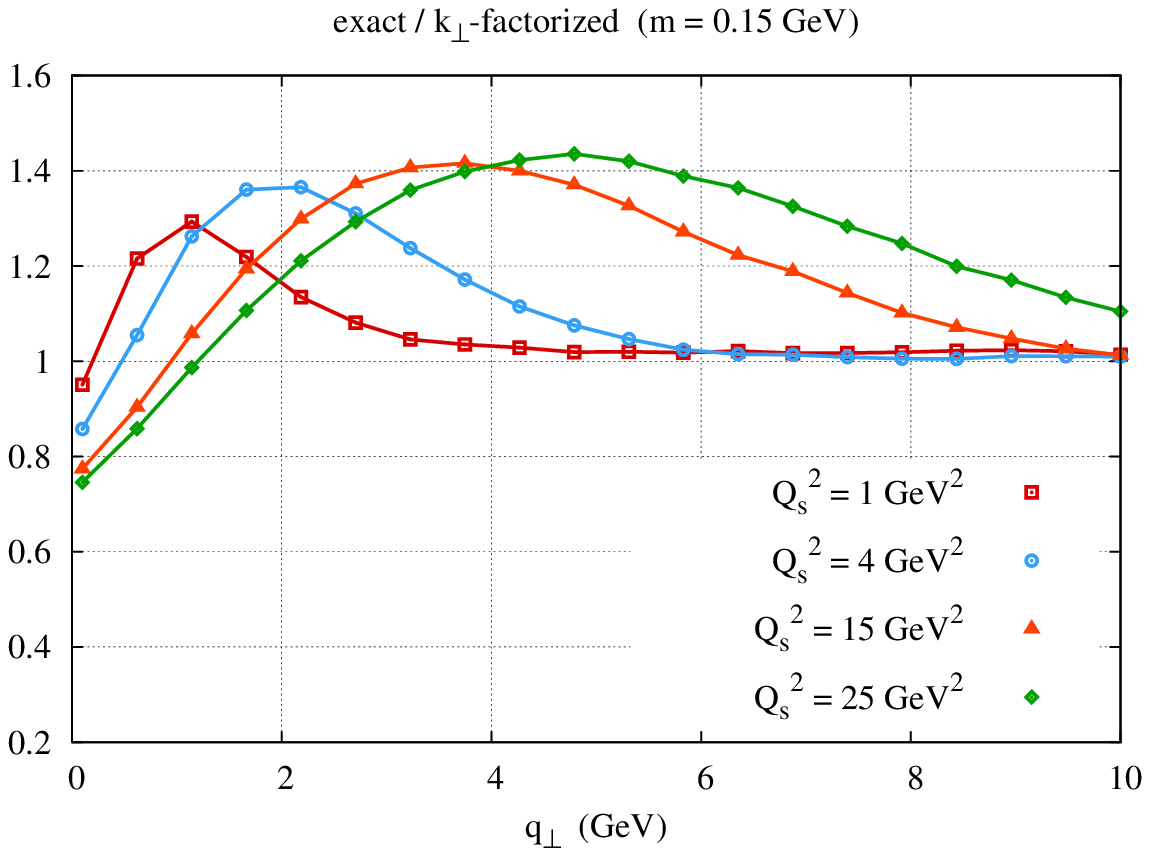}}
\vskip 1mm
\resizebox*{6.5cm}{!}{\includegraphics{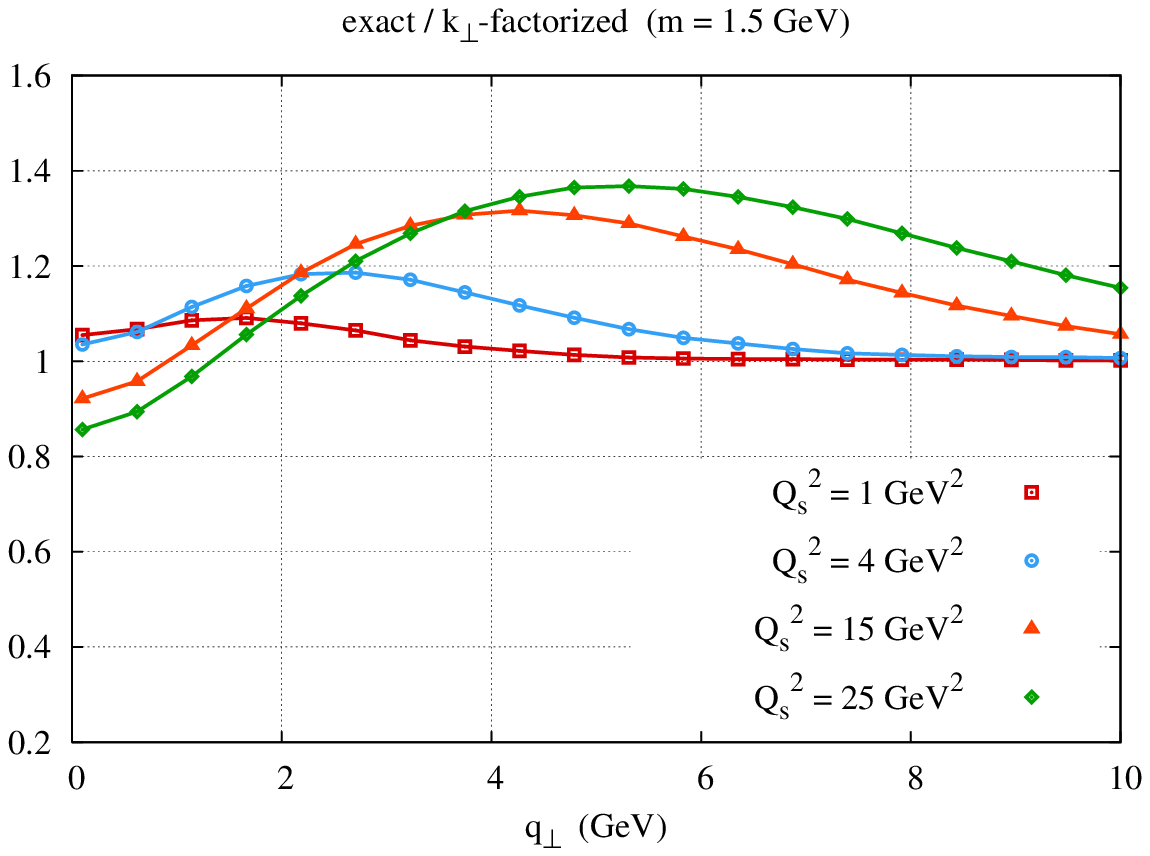}}
\vskip 1mm
\resizebox*{6.5cm}{!}{\includegraphics{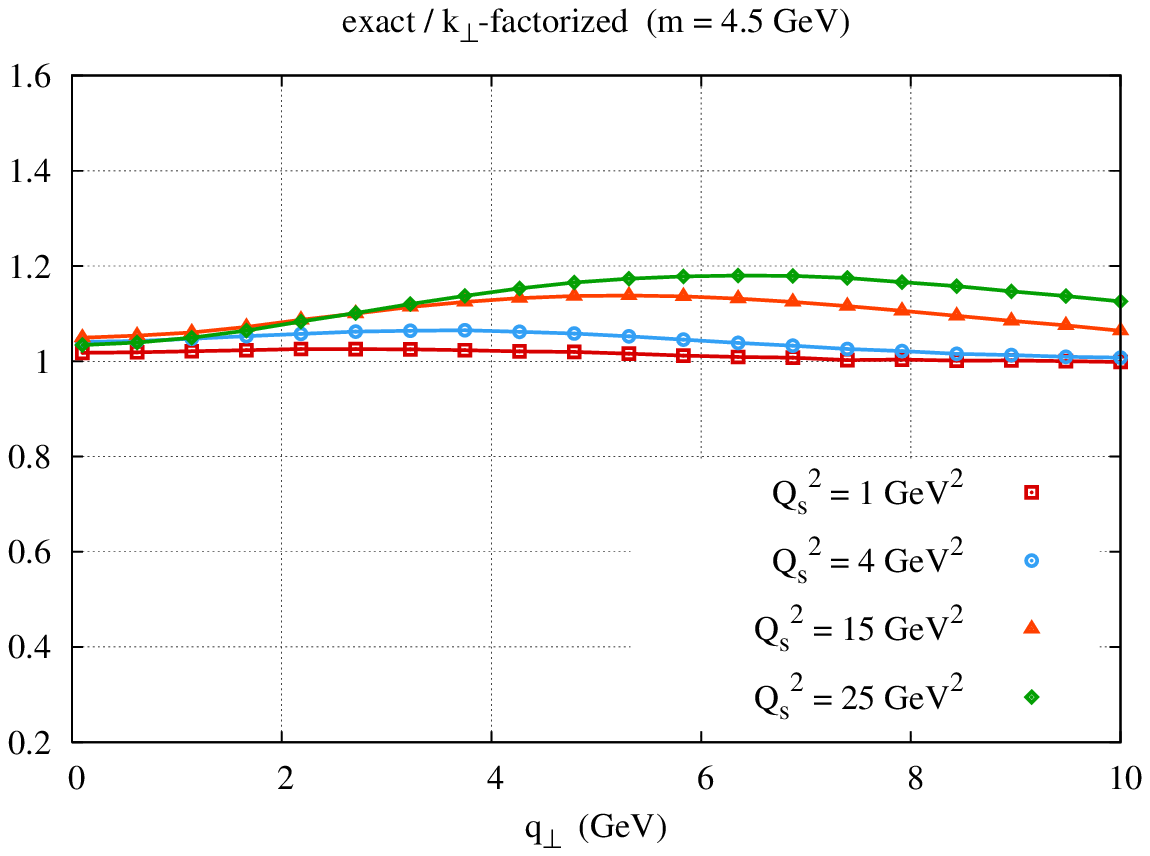}}
\end{center}
\caption{\label{fig:kt-breaking} Breaking of $k_\perp$-factorization
 in the MV model, respectively for strange, charm and bottom quark
 production.}
\end{figure}

We shall now discuss the violation of $k_\perp$-factorization in
cross-sections.  In Fig.~\ref{fig:kt-breaking}, we plot the ratio of the complete result to the
$k_\perp$ factorized result (obtained by using eq.~(\ref{eq:3p-kt-fact}) for the 3-point
function) as a function of the transverse momentum in the MV model, respectively for quark masses $m=0.15$~GeV,
$m=1.5$~GeV and $m=4.5$~GeV . The various curves in each figure
correspond to different values of $Q_s$. Extrapolating from saturation models of the HERA data (see ~\cite{CGC} for 
a discussion), one estimates $Q_s^2 = A^{1/3}\,(x_0/x)^\lambda$, where $x_0 = 3\cdot 10^{-4}$ and 
$\lambda\approx 0.3$. The values of $Q_s^2$ shown in Fig.~\ref{fig:kt-breaking} correspond to a 
wide range of $x$. The values of $Q_s^2=1$ GeV$^2$ and $4$ GeV$^2$ correspond to the central regions at 
RHIC and LHC respectively. Larger values of $15$ and $25$ GeV$^2$ correspond to very forward kinematics-between 
4 and 6 units of rapidity in the proton direction in future proton-Lead experiments at the LHC;  large values 
of $Q_s$ may also be accessed in proposed upgrades at RHIC. 

From the figure \ref{fig:kt-breaking}, one can clearly deduce the
following:

(1) The breaking of $k_\perp$-factorization is quite sensitive to the quark mass. 
The magnitude of the breaking is systematically smaller for heavier quarks because they are less 
sensitive to re-scattering effects than light quarks.
  
(2) The magnitude of the breaking for the heavier quarks becomes significant only at values of the 
saturation scale that correspond to forward rapidities. Isolating these effects from other uncertainities 
in these measurements is difficult but can be done in principle. One might have to look simultaneously at other final 
states, such as hard diffractive final states, that are also sensitive to three point functions. 
  
(3) The magnitude of its breaking is maximal for $q_\perp\sim Q_s$. One recovers $k_\perp$-factorization 
when the quark transverse momentum is much larger than $Q_s$.

(4) If $Q_s$ remains smaller than the quark transverse mass, the
  breaking of $k_\perp$-factorization enhances the
  cross-section because re-scatterings push a few more $Q\overline{Q}$ pairs above the
  pair production threshold. Conversely, if $Q_s$ is large relative to the quark mass, the cross-section 
  is reduced at small transverse momentum because multiple scatterings typically transfer
  large transverse momenta to the quark. 

\begin{figure}[htbp]
\begin{center}
\resizebox*{6.5cm}{!}{\includegraphics{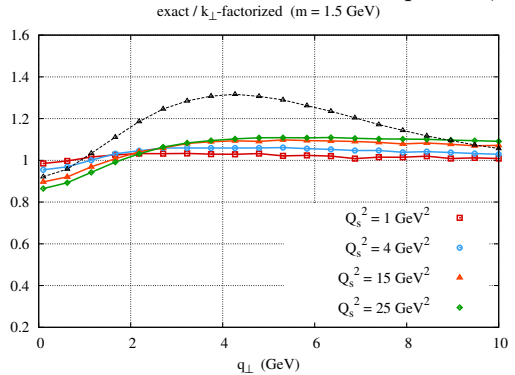}}
\end{center}
\caption{\label{fig:kt-breaking-c-asympt} Breaking of
$k_\perp$-factorization for charm quarks in the non-local Gaussian effective
model of \cite{IancuIM2}. The dashed line shows the result (for $Q_s^2 =15$ GeV$^2$) in the MV model for comparison.}
\end{figure}

As discussed previously, these conclusions in MV model are valid when parton densities are 
high but quantum evolution is not significant. Computing the same quantities in the GMF model 
provides a good idea of the direction and magnitude of the effects when small $x$ 
quantum evolution becomes significant. The
results in figure \ref{fig:kt-breaking-c-asympt} for the production of charm quarks, display the same trends outlined
previously.
The magnitude of the breaking of $k_\perp$-factorization is relatively smaller in this model. 
Similar results were obtained in our previous study ~\cite{BlaizGV1} of re-scattering
effects in gluon production. An intuitive way to understand this result is
to note that recombination/screening effects lead to a depletion in
the number of gluons available to re-scatter. For comparison, we also plot the result for charm quarks at   
$Q_s^2 =15$ GeV$^2$ in the MV model. We expect results of detailed RG computations to lie within the 
band outlined in Fig.~\ref{fig:kt-breaking-c-asympt}.

RV's research was supported in part by DOE Contract
No. DE-AC02-98CH10886 and the Alexander von Humboldt
foundation. He thanks the theory groups at Saclay,
Univ. of Bielefeld, the Univ. of Hamburg/DESY and ECT$\star$ for their
kind hospitality. HF's work is supported by the Grants-in-Aid for
Scientific Research (16740132).

\end{document}